\documentclass[preprint,showpacs,preprintnumbers,amsmath,amssymb]{revtex4}

\usepackage{epsfig}
\usepackage{graphics}
\usepackage{graphicx}
\usepackage{dcolumn}
\usepackage{bm}

\begin{document}

\preprint{APS/123-QED}

\title{Transition from anticipatory to lag synchronization via complete 
synchronization in time-delay systems}

\author{D.~V.~Senthilkumar}
\email{skumar@cnld.bdu.ac.in}
\author{M.~Lakshmanan}%
 \email{lakshman@cnld.bdu.ac.in}
\affiliation{%
Centre for Nonlinear Dynamics,Department of Physics,
Bharathidasan University, Tiruchirapalli - 620 024, India\\
}%

\date{\today}

\begin{abstract}
The existence of anticipatory, complete and lag synchronization in  a single
system having two different time-delays, that is feedback delay $\tau_1$ and
coupling delay $\tau_2$, is identified. The transition from anticipatory to
complete synchronization and from complete to lag synchronization as a function
of coupling delay $\tau_2$ with suitable stability condition is discussed. In
particular, it is shown that the stability condition is independent of the
delay times $\tau_1$ and $\tau_2$.  Consequently for a fixed set of parameters,
all the three types of  synchronizations can be realized. Further the emergence
of exact anticipatory/complete/lag synchronization  from the desynchronized
state via approximate synchronization, when one of the system parameters
($b_2$) is varied, is characterized by the minimum of similarity function and
the transition from on-off intermittency via periodic structure in laminar
phase distribution.
\end{abstract}

\pacs{05.45.Xt,05.45.Pq}
\maketitle

\section{\label{sec:level1}Introduction}
Synchronization phenomena dates back to the period of Huygens in 1665, when he
found that two very weakly coupled pendulum clocks, hanging from the same beam,
become phase synchronized~\cite{asp2001}.  Since the  early identification of
synchronization in chaotic oscillators~\cite{hfty1983, lmptlc1990,jfhlmp1995},
the phenomenon has attracted considerable research activity  in  different
areas of science~\cite{asp2001,mlkm1996,mlsr2003} and several  generalizations
and interesting applications have been developed.   Chaos synchronization
phenomenon is of interest not only from a theoretical point of view but also
has potential applications in diverse subjects such as as biological,
neurological, laser, chemical, electrical and fluid mechanical systems as well
as in secure communication, cryptography and so on.
~\cite{asp2001,hfty1983,lmptlc1990,jfhlmp1995,mlkm1996,mlsr2003,skhck1995,
lkup1995,csmgr1998,bbap1995}.  A recent review on the phenomenon of chaos
synchronization  can be found in the reference~\cite{sbjk2002}.  In recent
years, different kinds of synchronization have been identified: Complete (or
identical) synchronization~\cite{hfty1983,lmptlc1990}, generalized
synchronization ~\cite{nfr1995,lkup1996,rb1998}, phase
synchronization~\cite{mgrasp1996, ytycl1997}, lag
synchronization~\cite{mgrasp1997,srik2002,mzgww2002} and anticipatory
synchronization~\cite{huv2002,huv2001,cm2001}. For a critical discussion on the
interrelationship between various kinds of synchronization,  we may refer to
refs.~\cite{rblk2000,sblmp2001}.  Transition from one kind of synchronization
to the other, coexistence of different kinds of synchronization  in  time
series and also the nature of transition have also been studied
extensively~\cite{mgrasp1997,srik2002,mzgww2002,mzyw2003,alfr2001} in  coupled
chaotic systems.

One of the most important applications of chaos synchronization is secure
communication.  It is now an accepted fact that secure communication based on
simple low dimensional chaotic systems does not ensure sufficient level of
security, as the associated  chaotic attractors can be reconstructed with some
effort  and the hidden message can be retrieved by an
eavesdropper~\cite{lp1996}.  One way to overcome this problem is to consider
chaos synchronization in high dimensional systems having multiple positive
Lyapunov exponents.  This increases security by giving rise to much more
complex time series, which are apparently not vulnerable to the unmasking
procedures generally.  Recently chaotic  time-delay systems have been suggested as good
candidates for secure communication~\cite{mcmlg1977,jdf1982}, as the time-delay
systems are essentially  infinite dimensional in nature and are described by
delay differential equations, and that they can admit hyperchaotic attractors
with large number of positive Lyapunov exponents for suitable nonlinearity. 
Therefore the study of chaos synchronization in the time-delay systems is of
considerable practical significance. However, it should be noted that  one has
to be cautious due to the fact that even in time-delay systems with multiple
positive Lyapunov exponents  unmasking may be possible. Particularly, this is
so if any reconstruction of the dynamics of the system is achieved in some
appropriate space even for very high dimensional dynamics as demonstrated by
Zhou and Lai \cite{czchl1999} in the case of Mackey-Glass equation.

Time-delay is ubiquitous in many physical systems due to finite switching speed
of amplifiers, finite signal propagation time in biological networks, finite
chemical reaction times, memory effects and so on~\cite{mcmlg1977,
jdf1982,rjflg1993,ptkm1998}. In recent times, considerable work has been
carried out on the effect of time-delay in limit cycle
oscillators~\cite{dvras1998,dvras2000},  time-delay
feedback~\cite{jxukwc2003,jzypt2003}, networks with time-delay
coupling~\cite{mgeshs2003}, etc. Recently, we have shown  that even a single
scalar delay equation with piecewise linear function can exhibit hyperchaotic
behavior even for small values of time-delay~\cite{dvskml}. It is therefore of
importance to consider the synchronization of chaos in such scalar piecewise
linear delay differential systems with appropriate delay coupling. 
Interestingly, in the present work we find that in such a coupled system, one
can identify anticipatory, complete and lag synchronizations by simply tuning
the second time-delay parameter in the coupling, for a fixed set of system
parameters satisfying appropriate stability condition.  The results have  been
corroborated by the nature of similarity functions, and transition behavior
characterized by probability distribution of laminar phase during approximate
synchronization which precedes the exact synchronization when a system
parameter is varied.  We also wish to point out that to our knowledge such
transitions between all the three types of above synchronizations have not been
reported in nonhyperchaotic systems and it appears that the present type of
hyperchaotic systems are convenient tools to realize such transitions by tuning
the delay parameters suitably.

Specifically, in this paper we will consider chaos synchronization of two
single scalar  piecewise-linear time-delay systems studied in the
references~\cite{hlzh1996,ptkm1998,dvskml} with unidirectional coupling between
them and  having  two different time-delays: one in the coupling term and the
other in the individual systems, namely, feedback delay.  We have identified
the stability condition for synchronization following Krasovskii-Lyapunov
theory and  demonstrate that there exists transition between three different
kinds of synchronization, namely anticipatory, complete and lag
synchronizations, as a function of time-delay in the coupling.  To 
characterize the existence of anticipatory and lag synchronizations, we have
plotted the similarity function $S(\tau)$. We have also demonstrated that  when
the system parameter $b_2$ is varied,  the onset of exact
anticipatory/complete/lag synchronization from the desynchronized state is
preceded by a region of approximate synchronized state. We also show that the
later is characterized by a transition from on-off intermittency to a periodic
structure in the laminar phase distribution, as suggested in the work of Zhan
et al.~\cite{mzgww2002} for the case of lag synchronization. The plan of the 
paper is as follows. In sec.~II, we introduce the unidirectionally coupled
scalar time-delay system and identify the condition for stability of
synchronized states.  In  sec.~III, we point out the existence of anticipatory
synchronization when the strength of the coupling delay is less than feedback
delay, while in sec.~IV, complete synchronization is realized when the two
delays are equal.  Lag synchronization is shown to set in when the coupling
delay exceeds the feedback delay in sec.~V.  Finally in sec.~VI, we summarize
our results.

\section{Piecewise linear time-delay system and Stability condition for chaos
synchronization}
At first, we will introduce the single scalar time-delay system with
piecewise linearity and bring out its hyperchaotic nature for suitable
values of the system parameters. Then  the unidirectional delay coupling is introduced 
between two scalar systems and the stability condition for chaos synchronization
is derived.

\subsection{The scalar delay system}
We consider the following first order delay differential equation introduced
by Lu and He~\cite{hlzh1996} and discussed in detail by Thangavel et
al.~\cite{ptkm1998},
\begin{eqnarray}
\dot{x}(t)&=&-ax(t)+bf(x(t-\tau)),
\label{eq.onea}
\end{eqnarray}
where $a$ and $b$ are parameters, $\tau$ is the time-delay and $f$ is an
odd piecewise linear function defined as
\begin{eqnarray}
f(x)=
\left\{
\begin{array}{cc}
0,&  x \leq -4/3  \\
            -1.5x-2,&  -4/3 < x \leq -0.8 \\
            x,&    -0.8 < x \leq 0.8 \\              
            -1.5x+2,&   0.8 < x \leq 4/3 \\
            0,&  x > 4/3 \\ 
         \end{array} \right.
\label{eqoneb}
\end{eqnarray}
It is also of interest to consider additional forcing on the right
hand side of Eq.~(\ref{eq.onea}); however, this is not considered here.
The schematic form of ~(\ref{eqoneb}) is shown in Fig.~1.
Recently, we have reported ~\cite{dvskml}  that systems of the form
~(\ref{eq.onea}) exhibit hyperchaotic behavior for suitable parametric values.
For our present study, we find that for the choice of the parameters
$a=1.0, b=1.2$ and $\tau=25.0$ with the initial condition $x(t)=0.9,
t\in(-5,0)$, Eq.~(\ref{eq.onea}) exhibits hyperchaos.  The corresponding
pseudoattractor is shown in the Fig.~2.
The hyperchaotic nature of Eq.~(\ref{eq.onea}) is confirmed by
the existence of multiple positive Lyapunov exponents.  The first ten maximal
Lyapunov exponents for the parameters $a=1.0, b=1.2, x(t)=0.9, t\in(-5,0)$,
as a function of time-delay $\tau$ is shown in Fig.~3, which
are evaluated using the procedure suggested by J.~D.~Farmer~\cite{jdf1982}.

\subsection{Coupled system and the stability condition}  
Now let us consider the following unidirectionally coupled drive $x_1(t)$ and response
$x_2(t)$ systems with two different time-delays $\tau_1$ and $\tau_2$ as
feedback and coupling time-delays, respectively,
\begin{subequations}
\begin{eqnarray}
\dot{x_1}(t)&=&-ax_1(t)+b_{1}f(x_1(t-\tau_{1})),  \\
\dot{x_2}(t)&=&-ax_2(t)+b_{2}f(x_2(t-\tau_{1}))+b_{3}f(x_1(t-\tau_{2})),
\end{eqnarray}
\label{eq.one}
\end{subequations}
where $b_1, b_2$ and $b_3$ are constants, $a>0$, and $f(x)$ is of the same form
as in Eq.~(\ref{eqoneb}).

Now we can deduce the stability condition for synchronization of the two
time-delay systems Eqs.~(\ref{eq.one}a) and (\ref{eq.one}b) in the presence
of the delay coupling  $b_{3}f(x_1(t-\tau_{2}))$.  
The time evolution of the difference system with the state variable
$\Delta=x_{1\tau_2-\tau_1}-x_2$ (which corresponds to anticipatory
synchronization when $\tau_2 < \tau_1$, identical synchronization for $\tau_2 =
\tau_1$ and lag synchronization when $\tau_2 > \tau_1$), where 
$x_{1\tau_2-\tau_1} = x_{1}(t-(\tau_2-\tau_1))$, can be  written for
small values of $\Delta$ by using the evolution Eqs.~(3) as
\begin{align}
\dot{\Delta}=-a\Delta+(b_2+b_3-b_1)f(x_1(t-\tau_2))+b_2f^{\prime}
(x_1(t-\tau_2))\Delta_{\tau_1},\;\;\Delta_\tau=\Delta(t-\tau)  
\end{align}
In order to study the stability of the
synchronization manifold, we choose the parametric condition,
\begin{align}
b_1=b_2+b_3,
\label{eq.paracon}  
\end{align}
so that the evolution equation for the difference system $\Delta$ becomes
\begin{align}
\dot{\Delta}=-a\Delta+b_{2}f^\prime(x_1(t-\tau_{2}))\Delta_{\tau_1}.
\label{eq.difsys}
\end{align}
The synchronization manifold is locally attracting if the origin of this
equation is stable.  Following Krasovskii-Lyapunov functional approach [39,40],
 we define a positive
definite Lyapunov functional of the form
\begin{align}
V(t)=\frac{1}{2}\Delta^2+\mu\int_{-\tau_1}^0\Delta^2(t+\theta)d\theta,
\end{align}
where $\mu$  is an arbitrary positive parameter, $\mu>0$.  Note that $V(t)$
approaches zero as $\Delta \rightarrow 0$.

To estimate a sufficient condition for the stability of the solution $\Delta=0$,
we require the derivative of the functional $V(t)$ along the trajectory
of Eq.~(\ref{eq.difsys}),
\begin{align}
\frac{dV}{dt}=-a\Delta^2+b_2f^{\prime}(x_1(t-\tau_2))\Delta
\Delta_{\tau_1}+\mu\Delta^2-\mu\Delta_{\tau_1}^2,
\end{align}
to be negative.  The above equation can be rewritten as
\begin{align}
\frac{dV}{dt}&=-\mu\Delta^2\Gamma(X,\mu),
\end{align}
where $X=\Delta_{\tau_1}/\Delta$, $\Gamma=\bigr[\bigr((a-\mu)/\mu\bigl)
-\bigr(b_2f^{\prime}(x_1(t-\tau_2))/\mu\bigl)X+X^2\bigl]$.
In order to show that $\frac{dV}{dt}<0$ for all $\Delta$ and $\Delta_\tau$
and so for all $X$, it is sufficient to show that $\Gamma_{min}>0$.
One can easily check that the absolute minimum of $\Gamma$ occurs at
$X=\frac{1}{2\mu}b_2f^{\prime}(x_1(t-\tau_2))$ with 
$\Gamma_{min}=\bigr[4\mu(a-\mu)-b_2^2f^{\prime}
(x_1(t-\tau_2))^2\bigl]/4\mu^2$.
Consequently, we have the condition for stability as
\begin{align}
a>\frac{b_2^2}{4\mu}f^{\prime}(x_1(t-\tau_2))^2+\mu = \Phi(\mu).
\label{eq.ineq}
\end{align}
Again $\Phi(\mu)$ as a function of $\mu$ for a given $f^{\prime}(x)$ has an
absolute minimum at $\mu=(|b_2f^{\prime}(x_1(t-\tau_2))|)/2$ with 
$\Phi_{min}=|b_2f^{\prime}(x_1(t-\tau_2))|$.  Since $\Phi\ge\Phi_{min}=
|b_2f^{\prime}(x_1(t-\tau_2))|$, from the inequality (\ref{eq.ineq}), it turns out that
the sufficient condition for asymptotic stability is
\begin{align}
a>|b_2f^{\prime}(x_1(t-\tau_2))|
\label{eq.asystab}  
\end{align}
along with the condition (\ref{eq.paracon}) on the parameters $b_1,b_2$ and $b_3$.

Now from the form of the piecewise linear function $f(x)$ given by Eq.~(2),
we have,
\begin{align}
|f^{\prime}(x_1(t-\tau_2))|=
\left\{
\begin{array}{cc}
1.5,& 0.8\leq|x_1|\leq\frac{4}{3}\\
1.0,& |x_1|<0.8 \\
\end{array} \right.
\end{align}
Note that the region $|x_1|>4/3$ is outside the dynamics of the present system
(see Eq.~(\ref{eqoneb})). Consequently the stability condition
(\ref{eq.asystab}) becomes $a>1.5|b_2|>|b_2|$ along with the parametric
restriction $b_1=b_2+b_3$.

Thus one can take $a>|b_2|$ as a less stringent condition for (\ref{eq.asystab}) to
be valid, while
\begin{align}
a>1.5|b_2|, 
\label{eq.four}
\end{align}
as the most general condition specified by (\ref{eq.asystab}) for asymptotic
stability of the synchronized state $\Delta=0$.  The condition (\ref{eq.four})
indeed corresponds to the stability condition for exact anticipatory,
identical  as well as lag synchronizations for suitable values of the coupling
delay $\tau_2$. It may also be noted that the stability condition
(\ref{eq.four}) is independent of the both the delay parameters $\tau_1$ and
$\tau_2$. In the following, we will demonstrate the transition from
anticipatory to lag synchronization via complete synchronization as the
coupling delay $\tau_2$ is varied from $\tau_2 < \tau_1$ to $\tau_2 > \tau_1$,
subject to the stability condition~(\ref{eq.four}) with the parametric
restriction $b_1 = b_2+b_3$.  However, we also point out from detailed
numerical analysis that when the less general condition $1.5|b_2|>a>|b_2|$ is
satisfied, approximate synchronization (anticipatory/complete/lag) occurs.

\section{Anticipatory Synchronization for $\tau_2 < \tau_1$}
To start with, we first consider the transition to anticipatory
synchronization  in the coupled system~(\ref{eq.one}). We have fixed the value
of the feedback time-delay $\tau_1$ at $\tau_1=25.0$  while the other
parameters are fixed as $a=0.16, b_1=0.2, b_2=0.1, b_3=0.1$ and the time-delay
in the coupling $\tau_2$ is treated as the  control parameter.  With the above
mentioned  stability condition (\ref{eq.four})  and with the coupling delay
$\tau_2$ being less than the feedback delay $\tau_1$, one can observe the
transition to anticipatory synchronization.  The time trajectory plot is shown
in Fig.~4a depicting anticipatory synchronization, for the specific
value of $\tau_2=20.0$ with the anticipating time equal to that of difference
between feedback and the coupling delays, that is, $\tau=\tau_2 - \tau_1$.  The
time-shifted plot  Fig.~4b, $x_2(t-\tau)$ Vs $x_1(t)$, shows a
concentrated diagonal line confirming the existence of anticipatory
synchronization (We may note here that in all our numerical studies in this
paper we leave out sufficiently large number of transients, before presenting
our figures).

Sometime ago, Rosenbulm et al.~\cite{mgrasp1996} have introduced the notion of
similarity function  $S_l(\tau)$ for characterizing the lag synchronization
as a time averaged difference between the variables $x_1$ and $x_2$ (with mean
values being  subtracted) taken with the time shift $\tau$,
\begin{eqnarray}
S_l^2(\tau)=\frac{\langle[x_2(t+\tau)-x_1(t)]^2\rangle}
{[\langle x_{1}^2(t)\rangle\langle x_{2}^2(t)\rangle]^{1/2}},
\label{lag:sim}
\end{eqnarray}
where, $\langle x \rangle$ means time average over the variable $x$. If
the signals $x_1(t)$ and $x_2(t)$ are independent, the difference between them
is of the same order as the signals themselves.  If $x_1(t)$ = $x_2(t)$, as 
in the case of complete synchronization, the similarity function reaches a 
minimum  $S(\tau) = 0$ for $\tau = 0$.  But for the case of nonzero value
of time shift $\tau$, if $S_l(\tau) = 0$, then there exists a time shift $\tau$
between the two signals $x_1(t)$ and $x_2(t)$ such that $x_2(t+\tau) = x_1(t)$,
demonstrating lag synchronization.

In the present study, we have used the same similarity function $S_l(\tau)$ to
characterize anticipatory synchronization with negative time shift $-\tau$
instead of the positive time shift $\tau$ in Eq.~(\ref{lag:sim}). In other
words, one may define the similarity function for anticipatory synchronization
as
\begin{eqnarray}
S_a^2(\tau)=\frac{\langle[x_2(t-\tau)-x_1(t)]^2\rangle}
{[\langle x_{1}^2(t)\rangle\langle x_{2}^2(t)\rangle]^{1/2}}.
\label{anti:sim}
\end{eqnarray}
Then the minimum of $S_a(\tau)$,that is $S_a(\tau) = 0$, indicates that there
exists a time shift $-\tau$ between the two signals $x_1(t)$ and $x_2(t)$ such
that  $x_2(t-\tau) = x_1(t)$, demonstrating anticipatory synchronization. 
Fig.~5 shows the similarity function $S_a(\tau)$ as a  function of
the coupling delay $\tau_2$ for four different values of $b_2$, the parameter
whose value determines the stability condition given by Eq.~(\ref{eq.four}),
while satisfying the parametric condition $b_1=b_2+b_3$.  Curves 1 and 2 are
plotted for the values of $b_2 = 0.18 (>a=0.16>a/1.5)$ and $b_2 = 0.16
(=a>a/1.5)$, respectively, where the minimum values of $S_a(\tau)$ is found to
be greater than zero,  indicating that there is no exact time shift between the
two signals $x_1(t)$ and $x_2(t)$. Note that in the both cases the stringent
stability condition (\ref{eq.four}) and the less stringent condition $a>|b_2|$
are violated. Curve 3 corresponds to the value of $b_2=0.15$ (which is less
than $a$ but greater than $a/1.5$), where the minimum value of $S_a(\tau)$ is
almost zero, but not exactly zero (as may be seen in the inset of
Fig.~5), indicating an approximate anticipatory synchronization
$x_1(t)\approx x_2(t-\tau)$. On the other hand the curve 4 is plotted for the
value of $b_2 = 0.1 (<a/1.5)$, satisfying the  general stability criterion,
Eq.~(\ref{eq.four}).  It shows that the minimum of $S_a(\tau) = 0$, thereby
indicating that there exists an exact time shift between the two signals
demonstrating anticipatory synchronization.  The  anticipating time is found to
be equal to the difference between the coupling and feedback delay times, that
is, $\tau=\tau_2-\tau_1$. Note that $S_a(\tau) = 0$ for all values of $\tau_2
<\tau_1$, indicating anticipatory synchronization for a range of delay
coupling.  A further significance is that the anticipating time  $\tau =
|\tau_2-\tau_1|$ is an adjustable quantity as long as $\tau_2 <\tau_1$, which
can be tuned suitably to satisfy experimental situations.

Next, we show that the emergence of exact anticipatory synchronization is preceded by
a region of approximate anticipatory synchronization, which is associated with
the transition from on-off intermittency to a periodic structure in the laminar
phase distribution \cite{mzgww2002} as a function of the parameter $b_2$. First
we choose the value of $b_2$ as $b_2 = 0.17$ (with $b_1=0.2$ and $b_3=0.03$),
above the value of $a = 0.16$, such that the general stability criterion,
Eq.~(\ref{eq.four}),as well  the less stringent condition $a>|b_2|$ are
violated.

Fig.~6a shows the difference of $x_1(t)-x_2(t-\tau)$ Vs t,
exhibiting typical feature of on-off intermittency \cite{npsmh1994,jfhnp1994}
with the \emph{off} state near the laminar phase and the \emph{on} state
showing random burst. In Fig.~6b $x_1(t)$ is plotted against
$x_2(t-\tau)$, where the distribution is scattered around the diagonal. To analyze
the statistical feature associated with the irregular motion, we calculated
the distribution of laminar phases $\Lambda(t)$ with amplitude less than a threshold value
$\Delta=0.005$ as was done in the statistical analysis of intermittency
\cite{npsmh1994,jfhnp1994}, where the power law behavior of mean laminar
length is calculated as a function of control parameter.  A universal asymptotic
$-\frac{3}{2}$ power law distribution is observed in Fig.~6c,
which is quite typical for on-off intermittency.

Now, we choose the value of $b_2 = 0.15$, below the value of $a = 0.16$ so that
the less stringent condition $a>|b_2|$ is satisfied while the general stability
criterion Eq.~(\ref{eq.four}) is violated and we carry out the same  analysis
as above.  In Fig.~7a, the difference of  $x_1(t)-x_2(t-\tau)$ is
plotted against time $t$, which is more regular and is much smaller in
amplitude but not exactly zero, thereby implying an approximate anticipatory
synchronization $x_1(t) \approx x_1(t-\tau)$.  Fig.~7b shows the
plot of $x_1(t)$ Vs $x_2(t-\tau)$, where the distribution is localized entirely
on the diagonal, but not sharply on it.  Earlier we noted that for this case
the minimum of similarity function $S_a(\tau)$  (Curve 3, inset of
Fig.~5) is nearly zero, but not exactly zero. The distribution of
laminar phase $\Lambda(t)$ is plotted in Fig.~7c as for the
Fig.~6c. It shows a  periodic structure in the distribution of
laminar phase, where the peaks occur approximately at $t=nT, n=1,2,...$, where
$T$ is of the order of the period of the lowest periodic orbit of the
uncoupled system (\ref{eq.onea}). It should be remembered that the periodic
behavior is associated with the statistical analysis, while the signals remain
chaotic. Finally for the case $b_2=0.1 (<a/1.5)$, which satisfies the stringent
stability criterion~(\ref{eq.four}), and where the similarity function vanishes
exactly (Curve 4 in Fig.~5)), exact anticipatory synchronization
occurs as confirmed in Fig.~4.  Thus we find that the transition to
exact anticipatory synchronization precedes a region of approximate
anticipatory synchronization from desynchronized state as the parameter $b_2$
changes.  We have also
demonstrated that the emergence of this approximate anticipatory
synchronization from the desynchronized state is characterized by the
transition of on-off intermittency to periodic structure in the laminar phase
distribution.
\section{Complete Synchronization for $\tau_2 = \tau_1$}
Complete synchronization follows the anticipatory synchronization as the value
of the coupling time-delay $\tau_2$ equals the feedback time-delay $\tau_1$,
when $\tau_2$ is increased from a lower value.  With  $\tau_2 = \tau_1$, the
same stability criterion, Eq.~(\ref{eq.four}), holds good for this case of
complete synchronization as well with the same condition $b_1 = b_2+b_3$.

Fig.~8a shows the time trajectory plot of $x_1(t)$ and $x_2(t)$,
exhibiting synchronized evolution between them, which is also confirmed by the
entirely localized diagonal line of  $x_1(t)$ Vs $x_2(t)$ as shown  in
Fig.~8b. As in the case of anticipatory synchronization, we have 
found that the transition to complete synchronization precedes a region of
approximate complete synchronization ($x_1(t)\approx x_2(t)$) from the
desynchronized state as the parameter $b_2$ varies. Here also we have
identified that the emergence of  approximate complete synchronization for the
case $\tau_2=\tau_1$ is associated with a transition from on-off intermittency
to a periodic structure in the laminar phase distribution as a function of the
parameter $b_2$. In the next section we will discuss the existence of lag
synchronization for the values of $\tau_2$ greater than $\tau_1$.  

\section{Lag Synchronization for $\tau_2 > \tau_1$}
For coupling delay $\tau_2$ greater than feedback delay $\tau_1$, we find that
the system~(\ref{eq.one}) exhibits exact lag synchronization provided one satisfies
the stringent stability criterion (\ref{eq.four}), with the lag time equal to the
difference between the coupling and feedback delay times.

Fig.~9a shows the plot of $x_1(t)$ and $x_2(t)$ Vs time $t$, where
the response system lags the state of the drive system with constant lag time 
$\tau = |\tau_2-\tau_1|$. Fig.~9b shows the time-shifted plot
of $x_1(t)$ and $x_2(t+\tau)$. However, in the region of less stringent
stability condition, $1.5|b_2|<a<|b_2|$, approximate lag synchronization
occurs as in the cases of anticipatory and complete synchronizations.

We have also calculated the similarity function $S_l(\tau)$ from
Eq.~(\ref{lag:sim}) to characterize the lag synchronization. Fig.~10
shows the similarity function $S_l(\tau)$ Vs coupling delay $\tau_2$ for four
different values of $b_2$.  Curves 1 and 2 show the similarity function
$S_l(\tau)$ for the values of $b_2 = 0.18$ and $0.16$, respectively.  The
minimum of similarity function $S_l(\tau)$ occurs for  values of $S_l(\tau) >
0$ and hence there is a lack of exact lag time  between the drive and response
signals indicating asynchronization. Curve 3 corresponds to the value of
$b_2=0.15$ (which is less than $a$ but greater than $a/1.5$), where the minimum
values of $S_l(\tau)$ is almost zero, but not exactly zero  (as may be seen in
the inset of Fig.~10), so that $x_1(t)\approx x_2(t+\tau)$.   However
for the value of $b_3 = 0.1$, for which the general condition~(\ref{eq.four})
is satisfied, the minimum of similarity function  becomes exactly zero
(Curve 4) indicating that there is an exact time shift (Fig.~9) between drive
and  response signals $x_1(t)$ and $x_2(t)$,  respectively, confirming the
occurrence of lag synchronization.

We have also confirmed that as in the case of anticipatory synchronization, when
the parameter $b_2$ varies, the onset of exact lag synchronization is preceded
by a region of approximate lag synchronization, which is  characterized by a
transition from on-off intermittency of the desynchronized state to a periodic
structure in the laminar phase distribution. For the value of $b_2 = 0.17$
(which violates the stability condition~(\ref{eq.four}) as well as the less
stringent condition $a>|b_2|$), Fig.~11a shows the difference of
$x_1(t)-x_2(t+\tau)$ Vs time $t$, exhibiting a typical on-off intermittency. In
Fig.~11b, $x_1(t)$ is plotted against $x_2(t+\tau)$, where the
distribution is not concentrated along the diagonal.  In  Fig.~11a, the
laminar phase distribution $\Lambda(t)$ is characterized  by an exponential
$-\frac{3}{2}$ power law behavior as shown in the 

Fig.~11c.  In order to show that there is a transition from on-off
intermittency to periodic behavior in the laminar phase distribution
corresponding to approximate lag synchronization, we have changed the value of
$b_2$ from 0.17 to 0.15,  (so that the less stringent condition $a>|b_2|$ is
satisfied but not the general condition (\ref{eq.four})), and examined the
nature of laminar phase distribution $\Lambda(t)$.  The  difference between
$x_1(t)$ and  $x_2(t+\tau)$ is shown as a function of time $t$ in
Fig.~12a, where there is only a laminar phase present for the threshold
value $\Delta = 0.002$ without any intermittent burst. The corresponding
laminar phase distribution  $\Lambda(t)$ is again characterized by the periodic
structure as shown in Fig.~12c. As in the case of approximate
anticipatory synchronization, here also the peaks occur approximately at $t=nT,
n=1,2,...$, where $T$ is roughly of the order of the period of the lowest
periodic orbit of the uncoupled system (\ref{eq.onea}). Time-shifted plot
$x_1(t)$ Vs $x_2(t+\tau)$ is shown in the Fig.~12b,  where the
distribution is concentrated along but not exactly on the diagonal line
confirming the onset of approximate lag  synchronization. As noted previously
that for this case the minimum of similarity function $S_l(\tau)$ is nearly
zero but not exactly zero (Curve 3, inset of Fig.~10). Finally for
$b_2=0.1$, which satisfies the general stability criterion (\ref{eq.four}), we
have exact lag synchronization as demonstrated in Fig.~9 and
Fig.~10. Thus we find that as the parameter $b_2$ varies the transition
to exact lag synchronization precedes a region of approximate lag
synchronization from desynchronized state, where the later is characterized by
the transition from on-off intermittency to periodic structure in the laminar
phase distribution.
\section{Summary and conclusion}
In this paper, we have shown the existence of transition from anticipatory
synchronization  to lag synchronization through complete synchronization in a
single system of two coupled time-delay piecewise linear oscillators with
suitable  stability condition and with the second time-delay $\tau_2$ in the 
coupling as the only control parameter with all the other parameters being kept
fixed.  We have also plotted corresponding similarity functions to
characterize  both the anticipatory and lag synchronization as well as complete
synchronization. Further, when the parameter $b_2$ varies, we find that the
transition to exact anticipatory/complete/lag synchronization precedes a region
of approximate anticipatory/complete/lag synchronization from desynchronized
state, where the region of approximate   synchronization is characterized by
the transition from on-off intermittency to periodic structure in the laminar
phase distribution.

Further, we have observed that in the region where the stringent stability
condition (\ref{eq.four}) is satisfied, the minimum of  similarity function
$S_a(\tau)$  attains the value zero for all values of $\tau_2 < \tau_1$,
indicating that the exact anticipatory synchronization exists for a range of
coupling delay $\tau_2$ below $\tau_1$. However for approximate anticipatory
synchronization (in the region $1.5|b_2|>a>|b_2|$) the minimum of similarity
function takes the value $S_a(\tau)\approx 0$, but not exactly zero, for
$\tau_2 < \tau_1$.  Similarly lag synchronization also occurs for a range of
delay coupling  $\tau_2$ above $\tau_1$. Another interesting aspect is that
both the anticipating and lag time can be tuned to any desired value by
changing the value of coupling delay $\tau_2$. Consequently, the kind coupled
time-delay systems of the type discussed in this paper have considerable
physical relevance, particularly for  secure communication purposes.  We are
now investigating the existence of similar phenomena in other piecewise linear
time-delay systems, including the time-delay Chua and Murali-Lakshman-Chua
electronic circuits, the results of which will be published elsewhere.

\begin{acknowledgments}
This work has been supported by a Department of Science and Technology,
Government of India sponsored research project.
\end{acknowledgments}


\section{Figure Captions}

Fig1. The schematic form of the piecewise linear function
$f(x)$ given by Eq.~(\ref{eqoneb}).

Fig2. The hyperchaotic attractor of the system~(\ref{eq.onea})
 for the parameter values $a=1.0, b=1.2$ and $\tau=25.0$.

Fig3. The first ten maximal Lyapunov exponents
$\lambda_{max}$ of the scalar time-delay equation (\ref{eq.one}a) for the
parameter values $a=1.0, b=1.2,$  $\tau\in(2,29)$.

Fig4. Exact anticipatory synchronization for the parameter values
$a=0.16, b_1=0.2, b_2=0.1, b_3=0.1, \tau_1=25.0$ and $\tau_2=20.0$. (a) Time
series plot of $x_1(t)$ and $x_2(t)$, (b) Synchronization manifold between
$x_1(t)$ and  $x_2(t-\tau)$, $\tau=\tau_2-\tau_1$. The response $x_2(t)$
anticipates the drive $x_1(t)$ with a time shift of $\tau=5.0$.

Fig5. Similarity function $S_a(\tau)$ for different values
of $b_2$, the other system parameters are $a=0.16, b_1=0.2$ and 
$\tau_1=25.0$. (Curve~1: $b_2=0.18, b_3=0.02$, Curve~2: $b_2=0.16, b_3=0.04$,  
 Curve~3: $b_2=0.15, b_3=0.05$ and Curve~4: $b_2=0.1, b_3=0.1$).

Fig6. (a) The time series $x_1(t)-x_2(t-\tau)$ for 
$b_2=0.17$ and $b_3=0.03$ with all other parameters as in Fig.~4
(so that the stability condition is violated for anticipatory synchronization),
(b) Projection of $x_1(t)$ Vs $x_2(t-\tau)$ and (c) The statistical distribution
of laminar phase satisfying $-\frac{3}{2}$ power law scaling.

Fig7. (a) The time series $x_1(t)-x_2(t-\tau)$ for 
$b_2=0.15$ and $b_3=0.05$ with all other parameters fixed as in
Fig.~4 (so that the less stringent condition $a>|b_2|$ is satisfied
while (\ref{eq.four}) is violated), (b) Projection of $x_1(t)$ Vs $x_2(t-\tau)$
and (c) The statistical  distribution of laminar phase showing a periodic
structure.

Fig8. Exact complete synchronization for the parameter values $a=0.16,
b_1=0.2, b_2=0.1, b_3=0.1, \tau_1=25.0$ and $\tau_2=25.0$. Here the general
stability criterion (\ref{eq.four}) is satisfied. (a) Time series plot of
$x_1(t)$ and $x_2(t)$ and (b) Synchronization manifold between $x_1(t)$ and 
$x_2(t)$. The response $x_2(t)$ follows identically the drive $x_1(t)$ without
any time shift.

Fig9. Exact lag synchronization for the parameter values
$a=0.16, b_1=0.2, b_2=0.1, b_3=0.1, \tau_1=25.0$ and $\tau_2=30.0$. Here the
general stability criterion (\ref{eq.four}) is satisfied.(a) Time series plot
of $x_1(t)$ and $x_2(t)$, (b) Synchronization manifold between $x_1(t)$ and 
$x_2(t+\tau)$. The response $x_2(t)$ lags the drive $x_1(t)$ with a time shift
of $\tau=5.0$.

Fig.10 Similarity function $S_l(\tau)$ for different values
of $b_2$, the other system parameters are $a=0.16, b_1=0.2$ and 
$\tau_1=25.0$. (Curve~1: $b_2=0.18, b_3=0.02$, Curve~2: $b_2=0.16, b_3=0.04$ 
 Curve~3: $b_2=0.15, b_3=0.05$ and Curve~4: $b_2=0.1, b_3=0.1$).

Fig11. (a) The time series $x_1(t)-x_2(t+\tau)$ for  $b_2=0.17$
and $b_3=0.03$ with all other parameters as in Fig.~9 (so that the
 stability condition  is  violated), (b) Projection of $x_1(t)$ Vs
$x_2(t+\tau)$ and (c) The statistical distribution of laminar phase satisfying
$-\frac{3}{2}$ power law scaling.

Fig12. (a) The time series $x_1(t)-x_2(t+\tau)$ for  $b_2=0.15$
and $b_3=0.05$ so that the less stringent condition $a>|b_2|$ is satisfied
while (\ref{eq.four}) is violated, (b) Projection of $x_1(t)$ Vs $x_2(t+\tau)$
and (c) The statistical  distribution of laminar phase showing periodic
structure.

\end{document}